%% file: ms.tex
\documentclass[conference]{IEEEtran}
\IEEEoverridecommandlockouts

\usepackage{cite}
\usepackage{amsmath,amssymb,amsfonts}
\usepackage{algorithmic}
\usepackage{graphicx}
\usepackage{epstopdf}
\graphicspath{ {./images/} }
\usepackage{textcomp}
\usepackage{xcolor}
\usepackage[ruled,vlined,linesnumbered]{algorithm2e}
\usepackage{upgreek}

\usepackage{subcaption}

\usepackage{multirow}
\usepackage{array}
\newcolumntype{C}[1]{>{\centering\let\newline\\\arraybackslash\hspace{0pt}}m{#1}}
\newcolumntype{Q}[1]{>{\centering\arraybackslash\hspace{0pt}}m{#1}}

\let\oldnl\nl%
\newcommand\nonl{%
  \renewcommand{\nl}{\let\nl\oldnl}}%

\def\BibTeX{{\rm B\kern-.05em{\sc i\kern-.025em b}\kern-.08em
    T\kern-.1667em\lower.7ex\hbox{E}\kern-.125emX}}

\usepackage{xspace}
\usepackage{booktabs}
\usepackage{color}
\usepackage{colortbl}

\definecolor{a}{rgb}{0.9,0.95,0.95}%
\definecolor{b}{rgb}{0.99,0.99,0.99}%
\definecolor{debugcolor}{rgb}{0.0,0.0,0.0}
\begin{document}

\title{Payload-Mass-Aware Trajectory Planning on Multi-User Autonomous Unmanned Aerial Vehicles%
}

\author{
\IEEEauthorblockN{Vasileios Tsoutsouras}
\IEEEauthorblockA{\textit{University of Cambridge} \\
Cambridge, United Kingdom \\
vt298@cam.ac.uk}
\and
\IEEEauthorblockN{Joseph Story}
\IEEEauthorblockA{\textit{University of Cambridge} \\
Cambridge, United Kingdom \\
jdrs3@cam.ac.uk}
\and
\IEEEauthorblockN{Phillip Stanley-Marbell}
\IEEEauthorblockA{\textit{University of Cambridge} \\
Cambridge, United Kingdom \\
phillip.stanley-marbell@eng.cam.ac.uk}
}

\maketitle

\begin{abstract}
Future unmanned aerial vehicles (drones) will be shared by multiple users and 
will have to operate in conditions
where their fully-autonomous function is required.  Calculation of
a drone’s trajectory will be important but optimal trajectories cannot
be calculated unless mass and flight speed are taken into account.  This
article presents the case for on-drone trajectory planning in a
multi-user dynamic payload mass scenario, allowing a drone to
calculate its trajectory with no need for ground control communication.
We formulate and investigate on-drone trajectory planning under
variable payload mass and flight speed awareness, in cases where it is 
shared by multiple users or applications. We present efficient solutions
using a combination of heuristic and optimization algorithms.
To support this investigation, we present a new model for the power dissipation of drone propulsion 
as a function of speed and payload mass. 
We evaluate our proposed algorithmic solution on contemporary
embedded processors and demonstrate its capability 
to generate near-optimal trajectories with limited computational
overhead (less than 300 milliseconds on an ARM Cortex-A9 SoC).
\end{abstract}

\input{sections/intro.tex}
\input{sections/related.tex}
\input{sections/formulation.tex}
\input{sections/proposed.tex}
\input{sections/evaluation.tex}

\section{Conclusion}

This work focused on trajectory planning under payload mass and flight speed awareness,
calculated as an integral module of the on-drone software stack.
In out target application, multiple users issue requests for payload delivery 
in specific waypoints under specific deadlines, forming a deadline-miss minimization problem.
We examined a solution combining a heuristic and optimizer algorithm, which we evaluated 
against benchmarks of varying input parameters using a data-driven power model 
from traces derived from simulated and actual drones.
Our experiments on two embedded systems showed the ability
of our module to provide good solutions of more than 80\% achieved deadlines
in the most demanding benchmarks, within a limited time of less than 4 sec.

\label{sec:conclusion}

\bibliographystyle{plain}
\bibliography{sigproc}

\end{document}

%% file: sections/intro.tex
\section{Introduction}
\label{sec:intro}

Computing devices enhanced with flying capabilities are enabling a new class of applications whose functionality inherently involves motion through space by one or more systems.
Unmanned aerial vehicles (UAVs) are already being actively investigated by both industry and academia~\cite{boroujerdian2018mavbench} leading to the creation of new application domains such as emergency condition tracking and relief, surveillance, package delivery, infrastructure monitoring, and mobile edge computing~\cite{hassanalian2017classifications}.

Flying computing systems impose a new set of challenges and open research questions~\cite{hassanalian2017classifications}.
Compute tasks share the same limited energy budget with mechanical parts, requiring a careful balance between the two.
Efficient computation becomes important because more complex and sophisticated control algorithms can enhance the flying characteristics of the UAV~\cite{boroujerdian2018mavbench}. However, system designers
must trade multiple factors for power consumption of the UAV as it varies with payload and flying speed~\cite{dorling2016vehicle, baek2018battery, abeywickrama2018empirical, tseng2017flight, goss2017realistic}.

\subsection{Payload mass variation over trajectories}
Many use cases require the UAV to be at a given location by a target time, thus the choice of flight trajectory for a given application flight task is an important parameter. 
In this work, our target application domain is payload delivery due to its ability 
to support a wide range of services (e.g., parcels, medicine delivery in disasters, etc.). Payload delivery is inherently characterized by changes of the UAV's aggregate mass over its flight trajectory.
We investigate applications with multiple target delivery waypoints issued by different users or applications, e.g., different companies
offering package delivery services while sharing the same UAV.

\begin{figure}
	\centering
	\includegraphics[width=0.8\columnwidth]{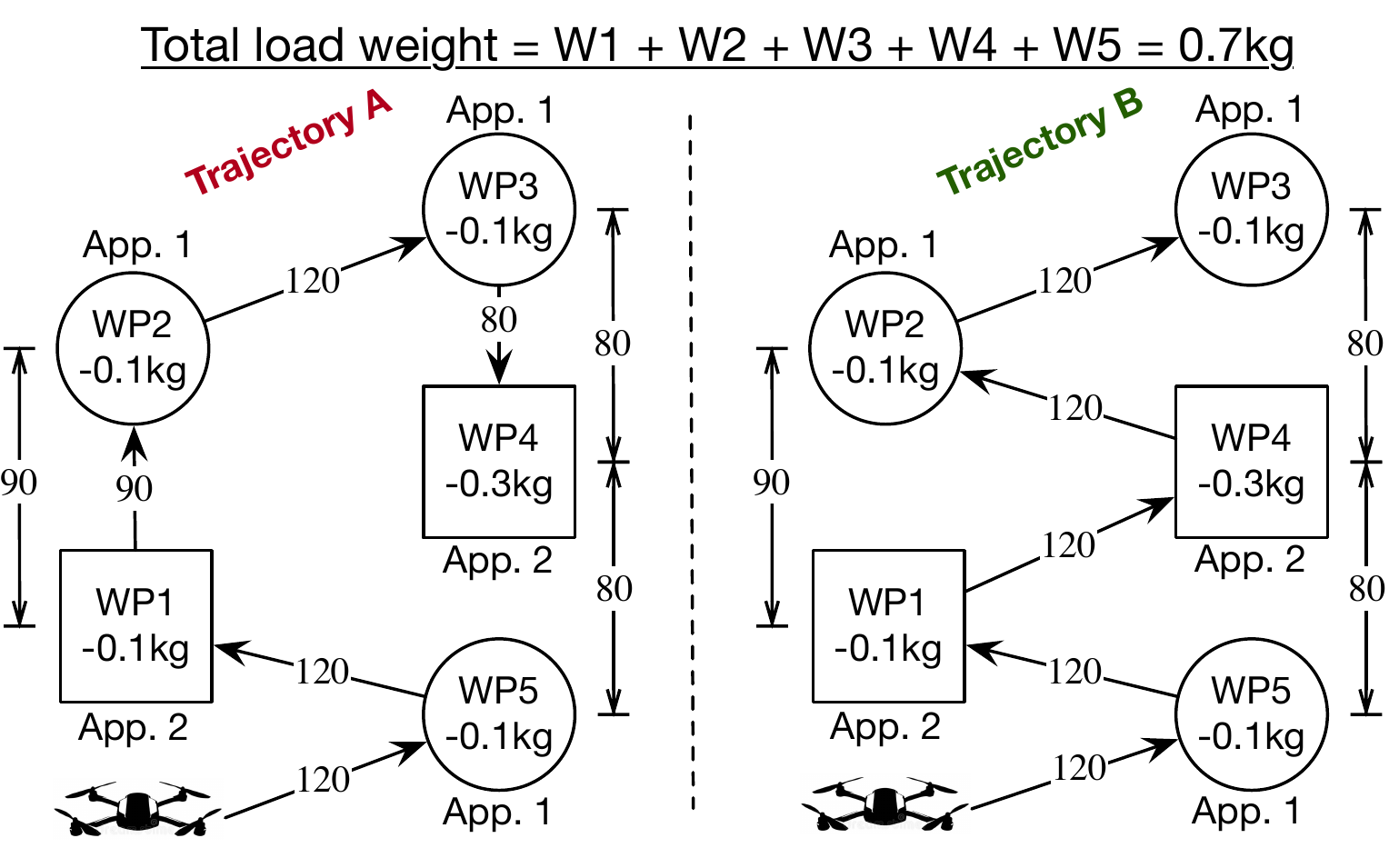}
	\caption{Payload-mass-aware trajectory designation for waypoints of Application 1 (circles) and Application 2 (squares).}
	\label{fig:trajectory-example}
\end{figure}

In this scenario, the definition of the optimum trajectory for visiting and unloading the payload is non-trivial, 
as Figure~\ref{fig:trajectory-example} shows.
In \textit{Trajectory A}, the UAV will fly a shorter distance but will be heavily loaded, since the
heaviest payload has to be delivered to \textit{Waypoint 4} (WP4), which is visited last.
On the other hand, in \textit{Trajectory B}, WP4 is visited earlier, so for the rest of the trajectory the
UAV flies with reduced weight.
As a result, the preferable trajectory route and speed cannot be established unless the relationship between
 energy used by a drone's propulsion system, payload, and flight speed is known.

We focus our analysis on quad-rotor UAVs or drones and opt for a fully-autonomous UAV
where trajectory calculation takes place on the on-board software stack.
Such decentralized decision making in embedded systems is an established
design approach which decouples the computing devices at the edge of the network
from their dependency to centralized resources~\cite{zhang2015cloud}. 
In a similar manner, our vision is that the navigation ability of the drone will be decoupled
from the connectivity hazards of ground station trajectory planning.

\subsection{Contributions}
Our main contributions are:
\begin{itemize}
\item \textbf{We present the formulation of a new multi-user payload
delivery application scenario} where multiple users can share the
same drone and issue requests for delivering payloads of a specified
mass at waypoints within specific deadlines. We formulate the
corresponding deadline-miss minimization problem under energy,
trajectory-waypoint, and mass constraints, taking into account the
variation of drone propulsion power consumption as a function of
payload mass and flight speed (Section~\ref{sec:formulation}).

\item \textbf{We propose a combination of heuristic and optimization
algorithms} for solving the deadline-miss minimization problem under
energy, trajectory-waypoint, and mass constraints.  Our proposed
design is fully-autonomous and trajectory calculation is performed
online on the drone (Section~\ref{sec:proposed}) using its
on-board computation resources, rather than offline on a server.

\item \textbf{We evaluate our solution for on-drone trajectory
planning} and show that it is low-overhead and effective
(Section~\ref{sec:evaluation}). The evaluation uses a power consumption
trace derived from an actual drone augmented with a data-driven
power model based on simulated drone flights.
\end{itemize}

%% file: sections/related.tex
\section{Related Research}
\label{sec:related}
Several related research efforts have focused on providing reliable models for predicting
the energy consumption of drones as a function of payload mass and flight speed.
Abeywickrama et al.~\cite{abeywickrama2018empirical} provide simple models for several parameters of the Intel 
\textit{Aero Ready to Fly Drone} and Goss et al. present a drone power consumption model~\cite{goss2017realistic}, by augmenting an analytical model of the drone's power consumption 
with data from the 3DR Solo drone. %

Baek at al.~\cite{baek2018battery} focus on a model for battery consumption including non-linear effects from increased
drawn current and depleted battery levels.
They propose a scheduling algorithm for payload delivery tasks, where the
drone delivers a single package and returns to a depot.
Tseng et al. \cite{tseng2017flight} provide a power model for drone power consumption under variable speed, 
payload, and different wind conditions, using the 3DR Solo drone.
Based on sampled data, they create a regression-based power model which they then use
offline to solve a path-planning problem where the drone must visit certain waypoints of interest
in the presence of recharging stations.

Routing and trajectory planning of UAVs has been a research focus for many years.
According to a recent survey, most approaches target scenarios of multiple UAVs and use a variety
of heuristics and optimization algorithms to solve the target problems, often modelled
as a Travelling Salesman's Problem variant~\cite{coutinho2018unmanned}.
One of the few approaches for the on-drone calculation of the trajectory is provided by Bandeira et al.~\cite{bandeira2015analysis}.
However, the problem they examined focuses on distance minimization under energy constraints and does not
take into account variable UAV drone payload mass and speed.

Dorling et al.~\cite{dorling2016vehicle} were among the first to identify the importance of a variable payload mass and speed model for drone power
consumption. They utilized a linear approximation to address the 
problem of delivery using multiple drones with the ability to execute multiple trips, but trajectories are calculated centrally and not on-drone. 	
Di Franco et al.~\cite{di2015energy} study path planning for image reconstruction of geographical
zones and propose an algorithm based on an energy model which takes into
account the speed of the drone as well as the resolution of the captured images. They construct their
model using data from the IRIS quadcopter, which is controlled by means of
the PX4 autopilot that we also used in our evaluation in Section~\ref{sec:evaluation}.

Cheng et al.~\cite{cheng2018formulations} provide a rigorous survey of drone routing. They suggest that most researchers
have focused on the problem of routing trucks with drones, where the drone delivers the payload in
the final destination in a single trip. They investigate the problem of routing a drone which can support
multiple trips, taking into account a non-linear model for drone power consumption
under variable payload mass and speed. They examine cases of multiple customers issuing waypoints
for multiple drones. Their solution is based on a \textit{branch and cut} algorithm and the
trajectory of all the drones is calculated in a centralized powerful server, not on a resource- and energy-constrained on-drone processor like we present in this work.

%% file: sections/formulation.tex
\section{Application and Problem Formulation}
\label{sec:formulation}

\begin{table}%
\caption{Notation and terminology.}
\begin{minipage}{4cm}
\def\arraystretch{1.0}\tabcolsep 2pt
\begin{tabular}{c|C{6.3cm}}
\toprule
\textbf{Notation} & \textbf{Description} \\
\hline
\rowcolor{a} $n$ & Number of waypoints in a trajectory, as requested by user applications. \\
\rowcolor{b} $\mathbb{C}$ & Set of coordinates of the $n$ input waypoints. \\
\rowcolor{a} $\mathbb{C_D}$ & Superset of $\mathbb{C}$ including the coordinates of start depot. \\
\rowcolor{b} $\mathbb{N_A} = \{1,...,n\}$ & Set of indices of the $n$ input waypoints. \\
\rowcolor{a} $\mathbb{N} = \{0,...,n+1\}$ & Superset of $N_A$ including index 0 for the start depot and index n+1 as its copy end depot. \\
\rowcolor{b} $W_d$ & Drone weight with no load. \\
\rowcolor{a} $V_{ij}$ & Velocity of drone as it travels from waypoint $i$ to $j$. \\
\rowcolor{b} $\mathbb{A}$ & Set unique user application identifiers (IDs). \\
\rowcolor{a} $w^{-}_{i}$ & Weight offloaded from the drone at waypoint $i$. \\
\rowcolor{b} $\mathbb{D}$ & Set of waypoint deadlines. \\
\rowcolor{a} $a^{i}_{j}$ & 1 if waypoint $i$ was requested by app. j, 0 otherwise. \\
\rowcolor{b} $x_{ij}$ & 1 if drone from waypoint $i$ to $j$, 0 otherwise. \\
\rowcolor{a} $l_{ij}$ & Straight distance between waypoints $i$ and $i$. \\
\rowcolor{b} $t_{i}$ & Time when waypoint $i$ is reached by drone. \\
\rowcolor{a} $q_{i}$ & Drone weight as it leaves waypoint $i$. \\
\rowcolor{b} $e_{0}, e_{min}$ & Initial available energy of the drone before departing the depot. Minimum acceptable remaining drone energy to ensure dependable flight.\\
\rowcolor{a} $V_{min}, V_{max}$ & Minimum and maximum acceptable drone velocity.\\
\rowcolor{b} $EC_{ij} $ & Energy consumption of drone, traveling through edge (i,j) with velocity $V_{ij}$ and $q_{i}$ weight.\\
\bottomrule
\end{tabular}
\label{tab:denotation}
\end{minipage}
\end{table}

\begin{table*}%
\caption{Problem formulation equations.}
\begin{tabular}{rm{9.7cm}}
\hline
minimize & \begin{equation} \label{eq:goal} N_{missed} = \sum_{\forall i \in \mathbb{N_A}} miss(t_{i}) \end{equation} \\[-4ex]
where & \begin{equation} \label{eq:goal2} miss(t_{i}) = \left\{ \begin{array}{lr} 0 & if \ t_{i} \leq d_{i}, i \in \mathbb{N_A}\\ 1 & otherwise \end{array} \right. \text{subject to:} \end{equation} \\[-4ex]
\textit{\color{gray}\small{All application waypoints visited exactly once}} & \begin{equation} \sum_{\forall j \in \mathbb{N_A} , i \neq j} x_{ij} = 1, \forall i \in \mathbb{N} \label{eq:cons_single_visit} \end{equation}\\[-4ex]
\textit{\color{gray}\small{Drone arriving at and leaving from same location}} & \begin{equation} \sum_{\forall j \in \mathbb{N}, i \neq j} x_{ij} -  \sum_{\forall j \in \mathbb{N}, i \neq j} x_{ji} = 0, \forall i \in \mathbb{N} \label{eq:cons_same_location} \end{equation}\\[-4ex]
\textit{\color{gray}\small{Subtour elimination constraint~\cite{oncan2009comparative}}} & \begin{equation} \sum_{\forall i,j \in S} x_{ij} \leq |S| - 1, \ S \subseteq \{2,...,n\}, \ 2 \leq |S| \leq n-1 \label{eq:cons_subtour} \end{equation}\\[-5ex]
\textit{\color{gray}\small{Arrival time at waypoint i, when departing from waypoint j}} & \begin{equation} t_i = \sum_{\forall j \in \mathbb{N}, i \neq j} x_{ji} \cdot (t_j + \frac{l_{ji}}{V_{ji}}), \forall i \in \mathbb{N_A} \label{eq:cons_time_calc} \end{equation}\\[-4ex]
\textit{\color{gray}\small{Calculated weight when departing waypoint i}} & \begin{equation} q_i = \sum_{\forall j \in \mathbb{N}, i \neq j} x_{ji} \cdot (q_j - w^{-}_{i}), \forall i \in \mathbb{N_A} \label{eq:cons_weight_calc} \end{equation}\\[-4ex]
\textit{\color{gray}\small{Energy consumption for reaching waypoint i from waypoint j}} & \begin{equation} EC_{ji} = \sum_{\forall j \in \mathbb{N}, i \neq j} x_{ji} \cdot E_d(q_{j}, V_{ji}, l_{ji}), \forall i \in \mathbb{N} \label{eq:cons_energy_calc} \end{equation}\\[-4ex]
\textit{\color{gray}\small{Decision variable value constraints}} & \begin{equation} x_{ij} \in \{0,1\}, q_{i} > 0 \ \forall i \in \mathbb{N} \label{eq:cons_values} \end{equation}\\[-5ex]
\textit{\color{gray}\small{Total initial drone weight when departing from depot}} & \begin{equation} q_0 = W_d + \sum_{\forall i \in \mathbb{N_A}} w^{-}_{i} \label{eq:cons_weight_0} \end{equation}\\[-4ex]
\textit{\color{gray}\small{Weight difference when departing from waypoint j}} & \begin{equation} \sum_{\forall i \in \mathbb{N}, i \neq j} x_{ij} \cdot q_{ij} -  \sum_{\forall k \in \mathbb{N}, k \neq j} x_{jk} \cdot q_{jk} = w^{-}_{j}, \forall j \in \mathbb{N_A} \label{eq:cons_weight_dif} \end{equation}\\[-4ex]
\textit{\color{gray}\small{Limit of total drone consumed energy}} & \begin{equation} \sum_{\forall i,j \in \mathbb{N}} x_{ij} \cdot EC_{ij} \leq e_0 - e_{min} \label{eq:cons_energy} \end{equation}\\[-5ex]
\textit{\color{gray}\small{Limits of drone velocity}} & \begin{equation} V_{min} \leq V_{ij} \leq V_{max} \label{eq:cons_velocity} \end{equation}\\[-5ex]
\textit{\color{gray}\small{Each waypoint belongs to a single application}} & \begin{equation} \sum_{\forall j \in \mathbb{N_A}} a^{i}_{j} = 1, \forall i \in \mathbb{A} \label{eq:cons_unique} \end{equation}\\[-6ex]
\end{tabular}
\label{tab:formulation}
\end{table*}

\textcolor{debugcolor}{
Most contemporary drones fly according to the requirements
of a single application or are remotely controlled by a single user.
Our target application model is based on the idea of multiple target
waypoints issued by multiple users or applications, whose interests might be
conflicting. The interests are expressed through a payload delivery
request within a certain deadline. Conflicts occur when not all interests can be satisfied in time, under the power constraints
of the drone. We assume on-drone
trajectory planning, freeing the drone from dependencies
on external ground control stations. Our driving example, is the one
of commercial payload delivery services, where requests are dictated by
multiple human users' needs. 
}

Let $n$ be the number of waypoints in a trajectory and let the
waypoints in a trajectory be indexed by $i>1$. Let $\mathbb{C}$ be
the set of coordinates of the $n$ waypoints and let $\mathbb{N_A}$
be the set of indices of the $n$ waypoints. Let $\mathbb{A}$ be
the set of unique application identifiers (IDs) that are relevant
to a given trajectory, indexed by $\alpha$. Each waypoint $i$ in $\mathbb{C}$
is issued by an application in $\mathbb{A}$ and at that waypoint a UAV may unload a weight $w^{-}_{i} \in \mathbb{W}$.
Picking up a load at a waypoint corresponds to unloading a weight
$w^{-}_{i} < 0$. Let $\mathbb{D}$ be a set of deadlines, one for
each waypoint in a trajectory, denoting the latest time by which
any of the applications $a \in \mathbb{A}$ require a payload of
weight $w^{-}_{i}$ to be unloaded. Let $\mathbb{C_D}$ be a superset of
$\mathbb{C}$ containing also the coordinates of the starting depot
station and let $\mathbb{N_A}$ be a superset of $\mathbb{N}$
containing index $0$ for the start depot and index $n+1$ for its
copy final depot.

The problem we wish to solve is to find a visiting order for the
$n$ waypoints in a trajectory given the $\mathbb{A}$, $\mathbb{W}$,
and the UAV's propulsion power dissipation model as a function of
flight speed and payload mass, such that the number of missed
deadlines across all applications and across all the waypoints is
minimized. The input of our target problem is a set $\mathbb{C}$
of $\vert\mathbb{C}\vert = n$ waypoints provided by user applications. %

The waypoints form a fully-connected graph $G(\mathbb{N}, \mathbb{E})$,
where $\mathbb{E}$ is the set of edges between coordinates $c \in
\mathbb{C}$, an edge $e_{ij} \in \mathbb{E} = \langle c_i, c_j\rangle$
is defined by a pair of coordinates, and the weights on the edges
correspond to the propulsion energy cost of travelling between
locations $c_i$ and $c_j$.  Because the cost per edge varies according
to the previously visited edges (e.g., payload is smaller after
unloading), the graph edge weights are asymmetric. \textit{The
target objective is that the drone leaves its start depot station
and returns to it, having first traversed all the application
waypoints, within the deadline of each one.}
We approach the problem as a deadline-miss minimization. Table~\ref{tab:formulation} presents the notation we use in the problem formulation.
Equation~\ref{eq:goal} defines the objective for minimizing the total number of the missed deadlines as requested by the user applications.
A miss is defined as a binary function, equal to 1 when the arrival time $t_i$ at waypoint $i \in \mathbb{N}$ is greater than the respective arrival deadline $d_i \in \mathbb{D}$.
Never visiting a target waypoint results also in a miss of the respective deadline.
Similar to the common formulations of a Travelling Salesman Problem~\cite{oncan2009comparative}, we use a binary variable $x_{ij}$ to represent the decision 
of whether the drone will fly from waypoint $i$ to waypoint $j$.
Using this variable, we formulate the trajectory constraints, i.e., each waypoint can be visited exactly once (Equation~\ref{eq:cons_single_visit}) and
a drone arriving in a waypoint $i$ is only allowed to depart from the same waypoint (Equation~\ref{eq:cons_same_location}). 
The combined effect of these equations is that each application waypoint is visited only once. 
Equation~\ref{eq:cons_subtour} formulates the Dantzig, Fulkerson, and Johnson (DFJ) sub-tour elimination constraint~\cite{oncan2009comparative}, which guards against
solutions which satisfy the constraints of Equation~\ref{eq:cons_single_visit} and Equation~\ref{eq:cons_same_location} but contain disjoint sub-tours of the input graph.

Assuming a constant flight speed $V_{ji}$ for travelling from waypoint $j$ to waypoint $i$, then the arrival time to waypoint $i$ is defined in Equation~\ref{eq:cons_time_calc} 
as the arrival time to waypoint $j$ and the ratio of $l_{ji}$ distance divided by the $V_{ji}$ speed. Note that $V_{ji}$ can differ for different pairs of $j$ and $i$.
Similarly, the weight when departing waypoint $i$ having arrived from waypoint $j$ is defined in Equation~\ref{eq:cons_weight_calc}, as the total drone weight when 
it departed waypoint $j$ minus the weight $w^{-}_{i}$ that the drone unloaded in waypoint $i$.
The energy that the drone consumes for flying from waypoint $j$ to waypoint $i$ is defined in Equation~\ref{eq:cons_energy_calc} as a function $E_d$
of the weight $q_j$ when departing waypoint $j$ and travelling through $e_{ji}$, the speed $V_{ji}$ and the distance $l_{ji}$.
In the general case, this function is non-linear and we present our methodology for deriving it in Section~\ref{subsec:power-model}.

Equation~\ref{eq:cons_values} defines $x_{ij}$ as a binary variable and constrains the variable $q_i$ to positive values.
Equation~\ref{eq:cons_weight_0} defines the drone weight when departing from the depot as the unloaded drone
weight added to the sum of the weights that need to be unloaded at all waypoints.
Equation~\ref{eq:cons_weight_dif} implies that the weight difference of entering and departing to and from waypoint 
$j$ is allowed to be only the unloaded weight $w^{-}_{j}$.
The most important drone constraint is its available energy for reliable flight. 
Equation~\ref{eq:cons_energy} constrains the available drone energy to be less or 
equal to the original drone energy $e_0$ minus the minimum required energy $e_{min}$ for the drone to fly.
Equation~\ref{eq:cons_velocity} constrains the drone flight velocity within acceptable limits.
Equation~\ref{eq:cons_unique} denotes that each waypoint is allowed to belong to only one application.
The key parameters of the system model are summarized in Table~\ref{tab:denotation}.
Our on-drone trajectory planning algorithm is responsible for designating the values of variables 
$x_{ij}$ and $V_{i,j}$ for each $i,j\in N$ so that deadlines misses are minimized under
all the constraints.

%% file: sections/proposed.tex
\section{Proposed solution}
\label{sec:proposed}
Drones have complex software stacks which comprise
multiple modules that need to be efficiently
executed at run-time. While some drone users opt for manual
control (e.g., via a phone app), many drones require autopilot software~\cite{di2015energy}.
These autopilots are designed to control the drone flight in order to
effectively execute a pre-loaded flight plan with pre-specified waypoints
and flying speed, set by a ground control software. 

\begin{figure}[t]%
	\centering
    \includegraphics[width=\columnwidth]{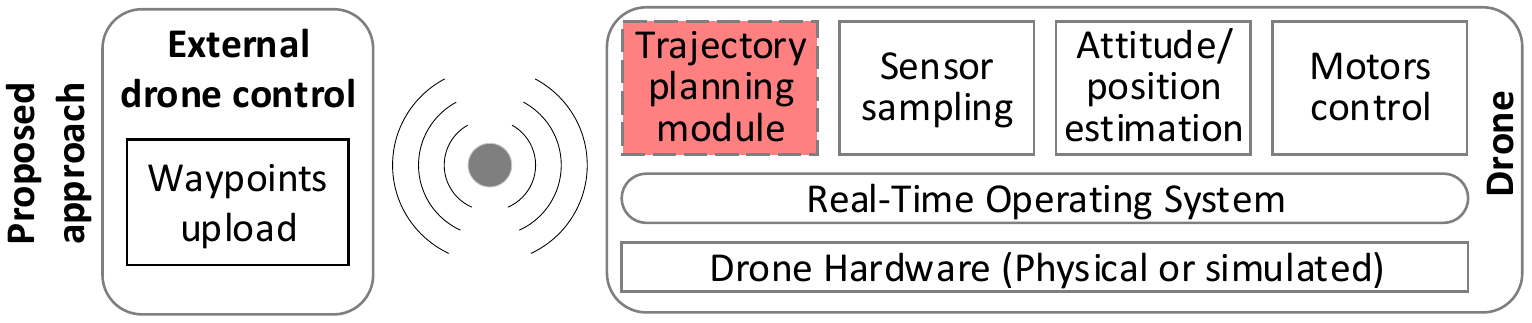}
    \caption{
    \textcolor{debugcolor}{On-drone software stack including our proposed trajectory planning module shaded dark.
    External control is used only for initial target waypoints upload.}}
    \label{fig:proposed-approach}
\end{figure}

In our proposal, illustrated in Figure~\ref{fig:proposed-approach}, 
drones can still communicate with a base station, but
this is only needed for uploading waypoints and related information
such as the issuing application, deadline, etc.
Uploading takes place before the drone leaves its starting depot and 
the trajectory-planning algorithm runs on the drone. The algorithm 
takes into account our problem formulation (Section~\ref{sec:formulation}) 
and is dynamically re-evaluated when the request for visiting new waypoints arises.
The algorithm determines the trajectory and flying speed of the drone, while
the rest of the existing drone control modules are responsible for
the trajectory execution.

These design choices remove the dependence of the drone on a ground control
station and allow for its trajectory to be recalculated when the drone
is far from the base station. The recalculation takes into account the
current energy capacity of the drone and its individual propulsion power
dissipation model, thus enabling the flight trajectory
to be tailored to the power management policy that the drone
management software intends to apply. This autonomy is achieved at the expense of a 
possibility of a sub-optimal trajectory plan due to the limited computational capabilities 
of the computing subsystems in most drones.

\subsection{Examined algorithms}
\label{subsec:algorithms}
We examine both heuristic and optimizer algorithms.
We design a heuristic, which builds a trajectory by greedily choosing the
nearer target waypoints with respect to the current drone position. 
We quantize the the set of flight speeds and evaluate the greedy algorithm for these quantized speed levels.

We also examine simulated annealing (SA), a well-known iterative optimizer, which has
already been used for drone routing~\cite{dorling2016vehicle}. 
We use the greedy heuristic solution as the input of the SA optimizer.
The latter is then able to fine-tune parameters like speed per
edge of the solution graph and thus enhance the solution quality.
Moreover, its light computational requirements and available control
parameters, enable the dynamic configuration of the intensity of solution searching.
Our experiments show that the effectiveness of the SA search is maximized
when combined with our greedy heuristic.

\begin{algorithm}[tb]
\footnotesize
\KwData{Waypoints coordinates, offloading payloads and deadlines}
\SetKwFunction{OnUAV}{Trajectory calculation}
  /* \textit{Initialization} */  
  
  loadPowerModel(); /* \textit{Load UAV-specific power model} */
  
  \While{true}{
	
	/* \textit{Blocking wait for input trajectory data} */	
	
	uploadedWaypoints[] = waitForRemoteUploadQueue();
	
	/* \textit{Get energy levels before trajectory calculation} */		
	
	remainingEnergy = checkBatteryStatus();
	
	/* \textit{Invoke Trajectory planning} */	
	
	cTrajectory = optimizeTrajectory(uploadedWaypoints[], remainingEnergy);
	
	/* \textit{Store/Queue trajectory for position controller} */	
	
	queueCalcTrajectory(cTrajectory, posContrQueue);
  }
  \caption{On drone trajectory planning module}
  \label{algo-on-drone}
\end{algorithm}

\begin{figure}[b]%
    \centering
    \includegraphics[width=1.0\columnwidth]{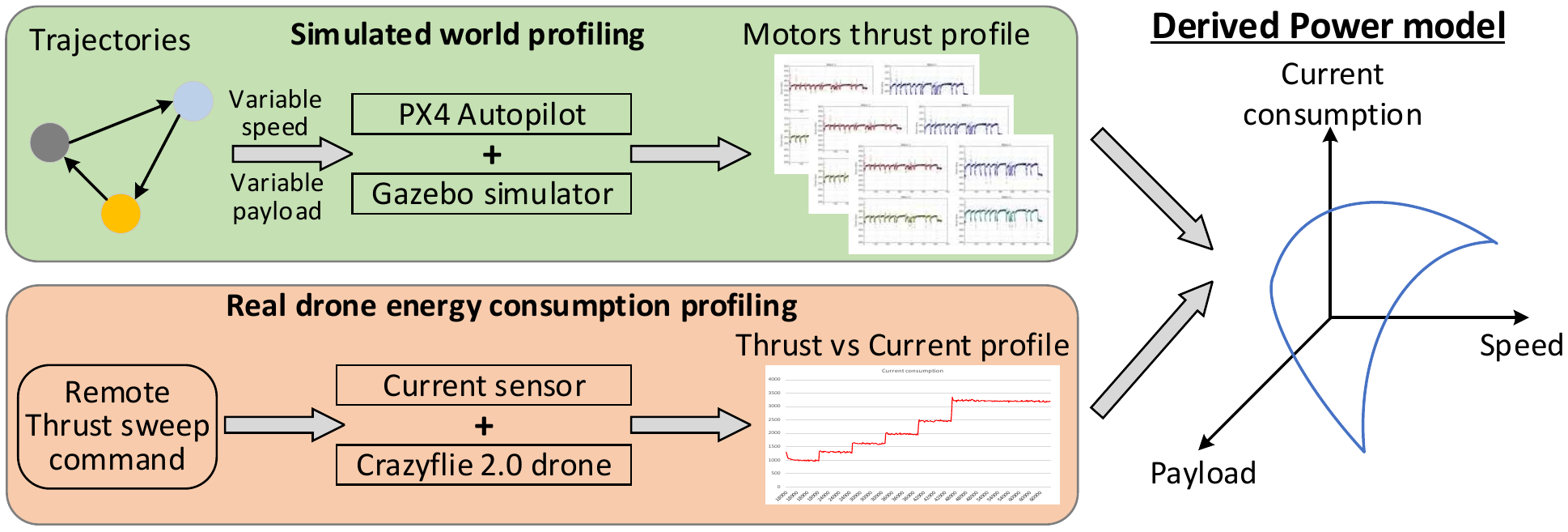}
    \caption{Drone power consumption modeling infrastructure.}
    \label{fig:power-model}
\end{figure}

Algorithm~\ref{algo-on-drone} presents the trajectory planning method we propose.
Following the approach of many autopilot stacks~\cite{meier2015px4, giernacki2017crazyflie}, 
different modules communicate through queues.
The initialization of the trajectory planning module corresponds to the
loading of the drone-specific propulsion power dissipation profile,
which is essential for the estimation of the energy requirements of a
candidate trajectory.
The module is then activated and blocks in its input queue, waiting for
a new set of input waypoints.
Upon this reception, it unblocks, updates its knowledge on remaining energy
and executes the trajectory planning algorithm.
The evaluated trajectory is then queued for execution by the position control module.

\begin{figure*}
       \makebox[\linewidth]{
       \centering   
       \begin{subfigure}[b]{0.33\linewidth}
       			\centering
                \includegraphics[width=\textwidth]{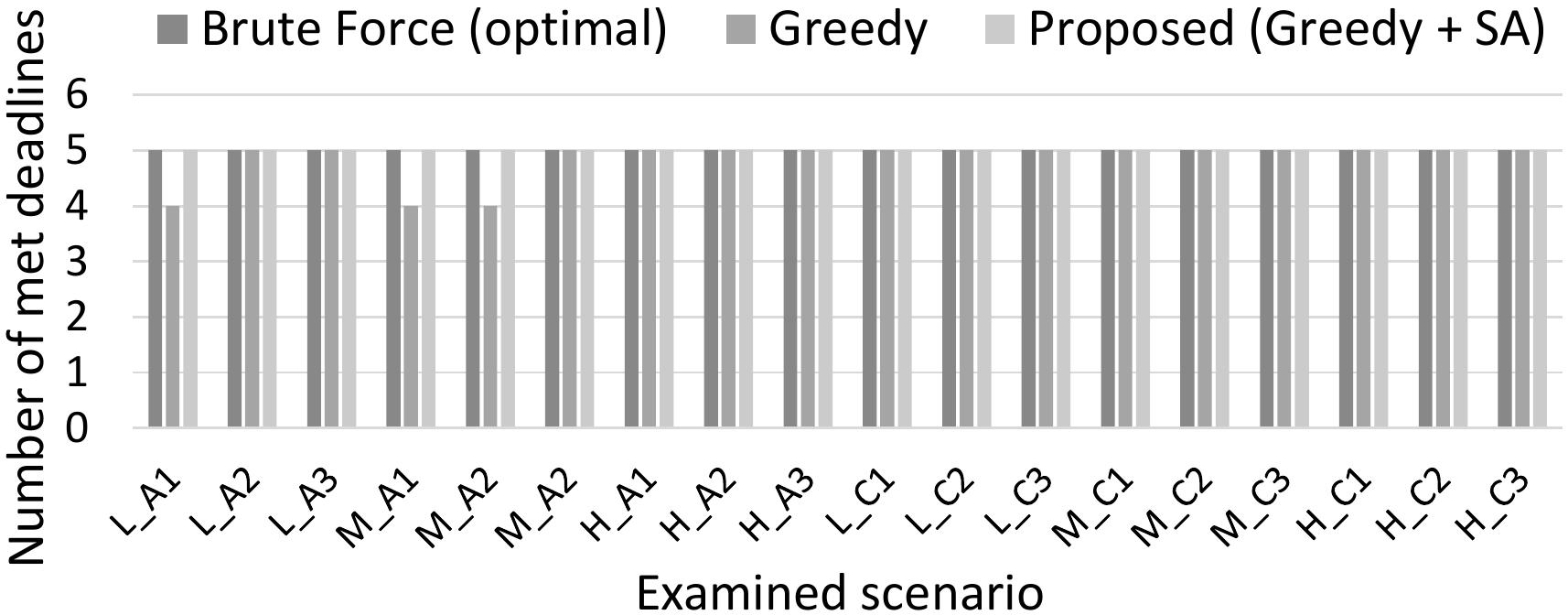}
                \caption{Benchmarks of 5 waypoints.}
                \label{fig:met-deadlines-5Wps}
        \end{subfigure}
        ~
		\begin{subfigure}[b]{0.33\linewidth}
       			\centering
                \includegraphics[width=\textwidth]{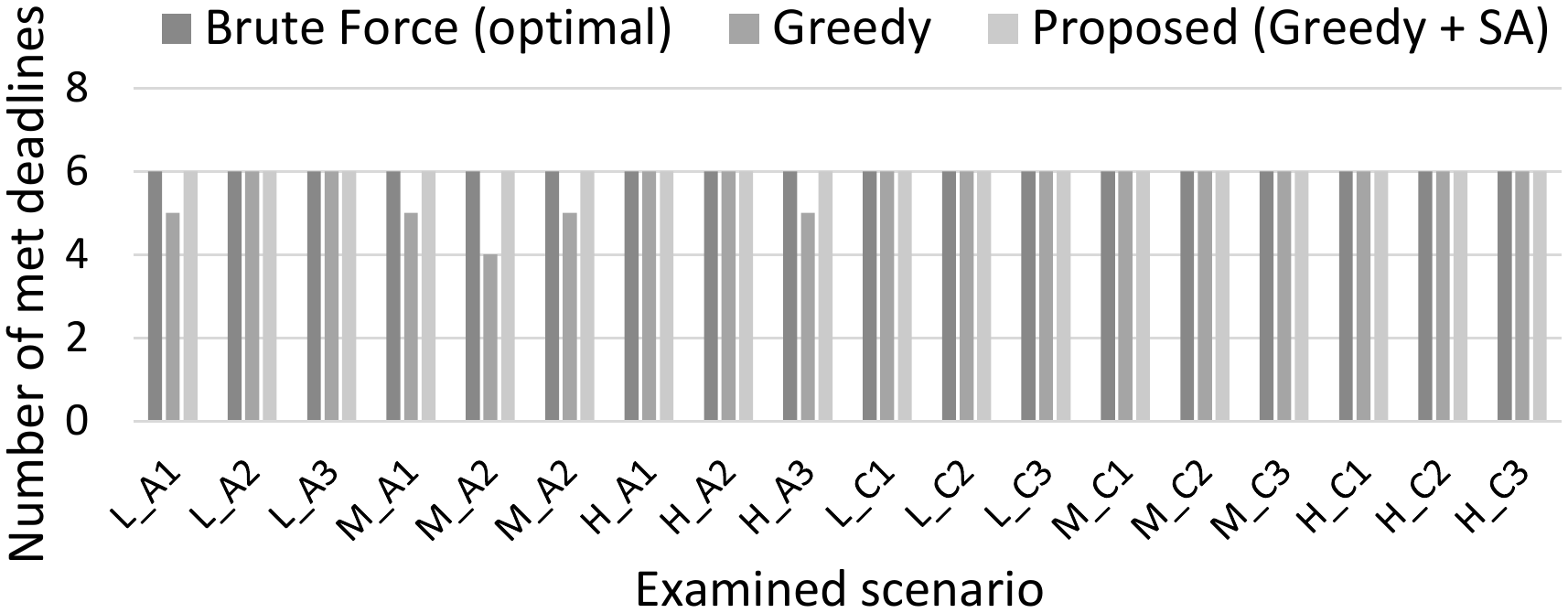}
                \caption{Benchmarks of 6 waypoints.}
                \label{fig:met-deadlines-6Wps}
        \end{subfigure}
        ~
        \begin{subfigure}[b]{0.33\linewidth}
       			\centering
                \includegraphics[width=\textwidth]{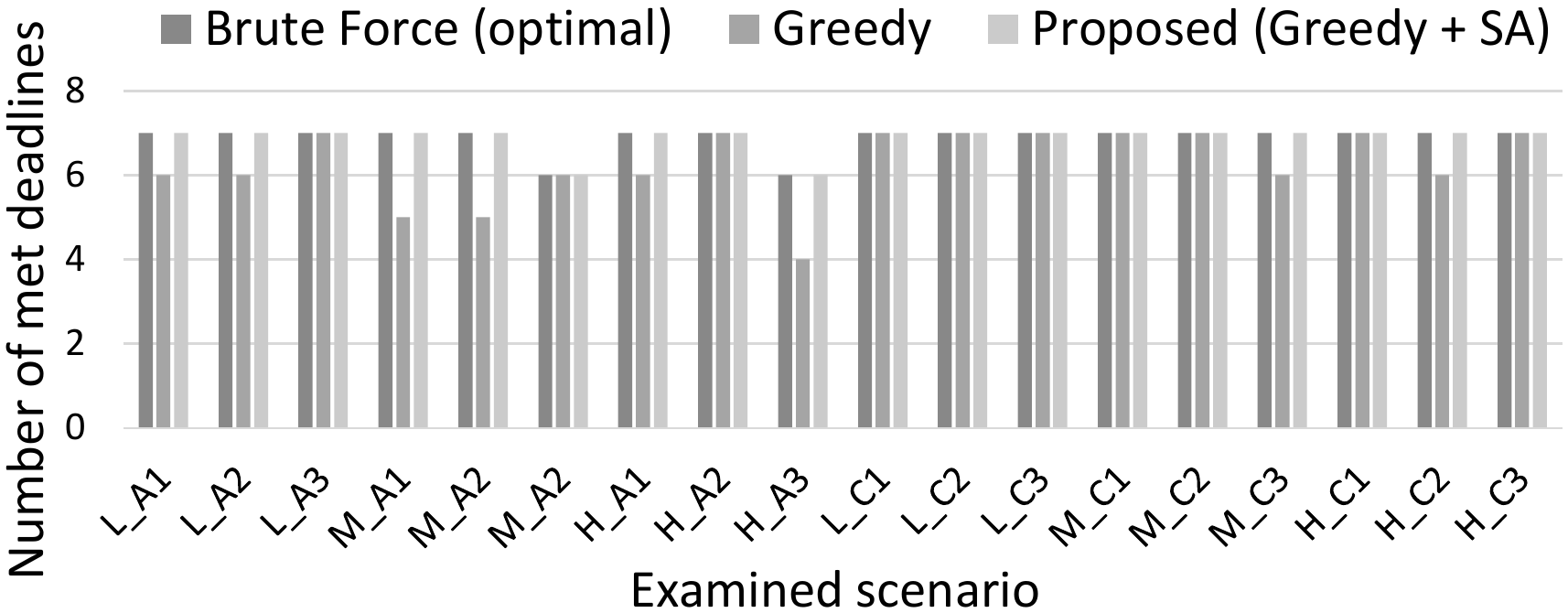}
                \caption{Benchmarks of 7 waypoints.}
                \label{fig:met-deadlines-7Wps}
        \end{subfigure}
        }
        ~\\
        \makebox[\linewidth]{
       \centering   
       \begin{subfigure}[b]{0.33\linewidth}
       			\centering
                \includegraphics[width=\textwidth]{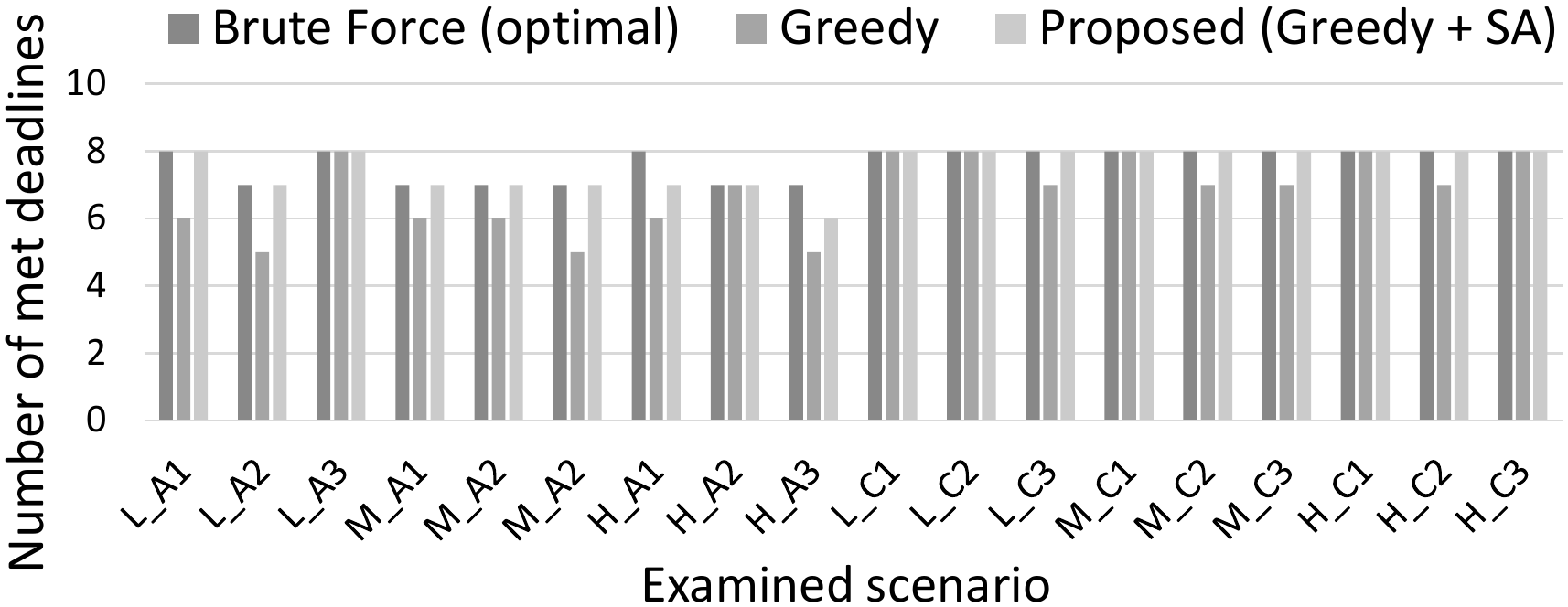}
                \caption{Benchmarks of 8 waypoints.}
                \label{fig:met-deadlines-8Wps}
        \end{subfigure}
        ~
		\begin{subfigure}[b]{0.33\linewidth}
       			\centering
                \includegraphics[width=\textwidth]{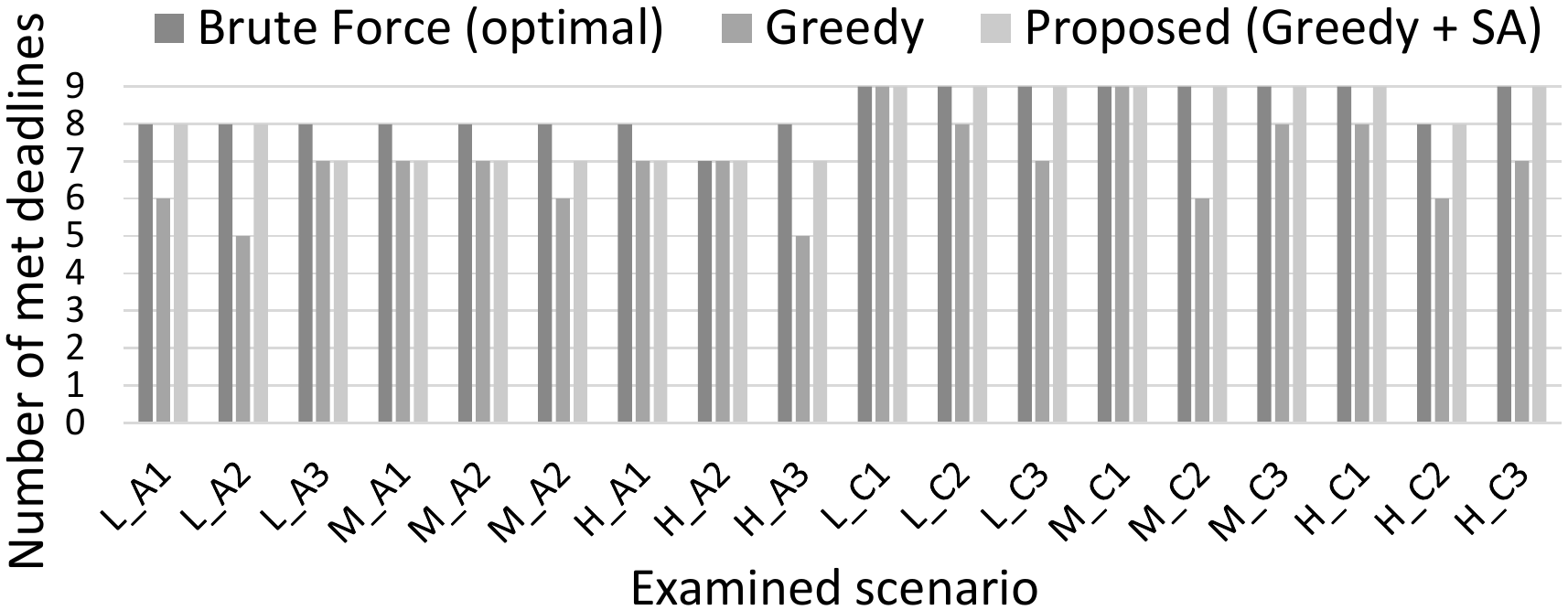}
                \caption{Benchmarks of 9 waypoints.}
                \label{fig:met-deadlines-9Wps}
        \end{subfigure}
        ~
        \begin{subfigure}[b]{0.33\linewidth}
       			\centering
                \includegraphics[width=\textwidth]{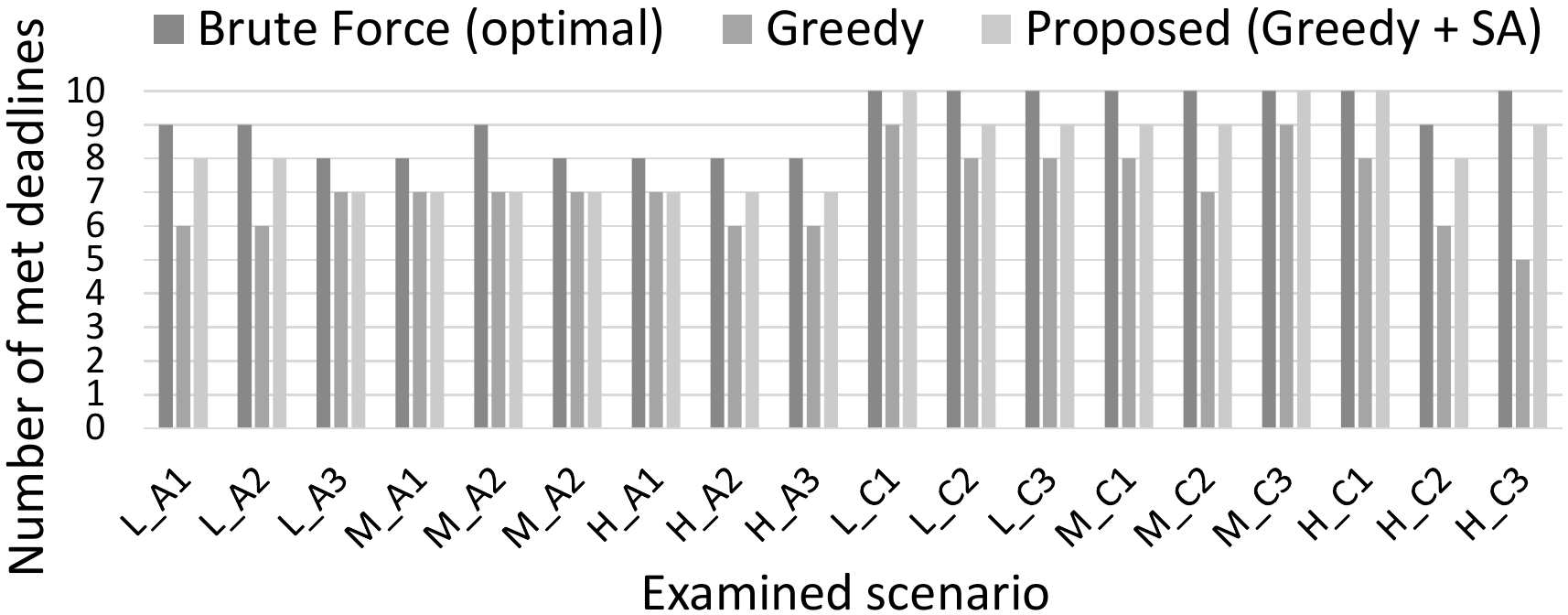}
                \caption{Benchmarks of 10 waypoints.}
                \label{fig:met-deadlines-10Wps}
        \end{subfigure}
        }
        \caption{Comparative analysis of solutions expressed via the number of met
        deadlines for each of the examined benchmarks.
		}
        \label{fig:met-deadlines-all}
\end{figure*}

\subsection{Analysis infrastructure --- power modeling}
\label{subsec:power-model}

Estimating the propulsion power dissipation
of a drone as a function of payload mass and flight speed is essential to the evaluation of our approach.
We used a simulation infrastructure
to profile different configurations of drone models for variable payload mass and flying speed.
\textit{We augment the simulation with actual power measurements of the current drawn by an actual drone (Crazyflie 2.0) 
to map the simulation trace to a propulsion power dissipation estimate.}

\begin{figure}[b]
       \makebox[\linewidth]{
       \centering   
       \begin{subfigure}[b]{0.40\linewidth}
       			\centering
                \includegraphics[width=\textwidth]{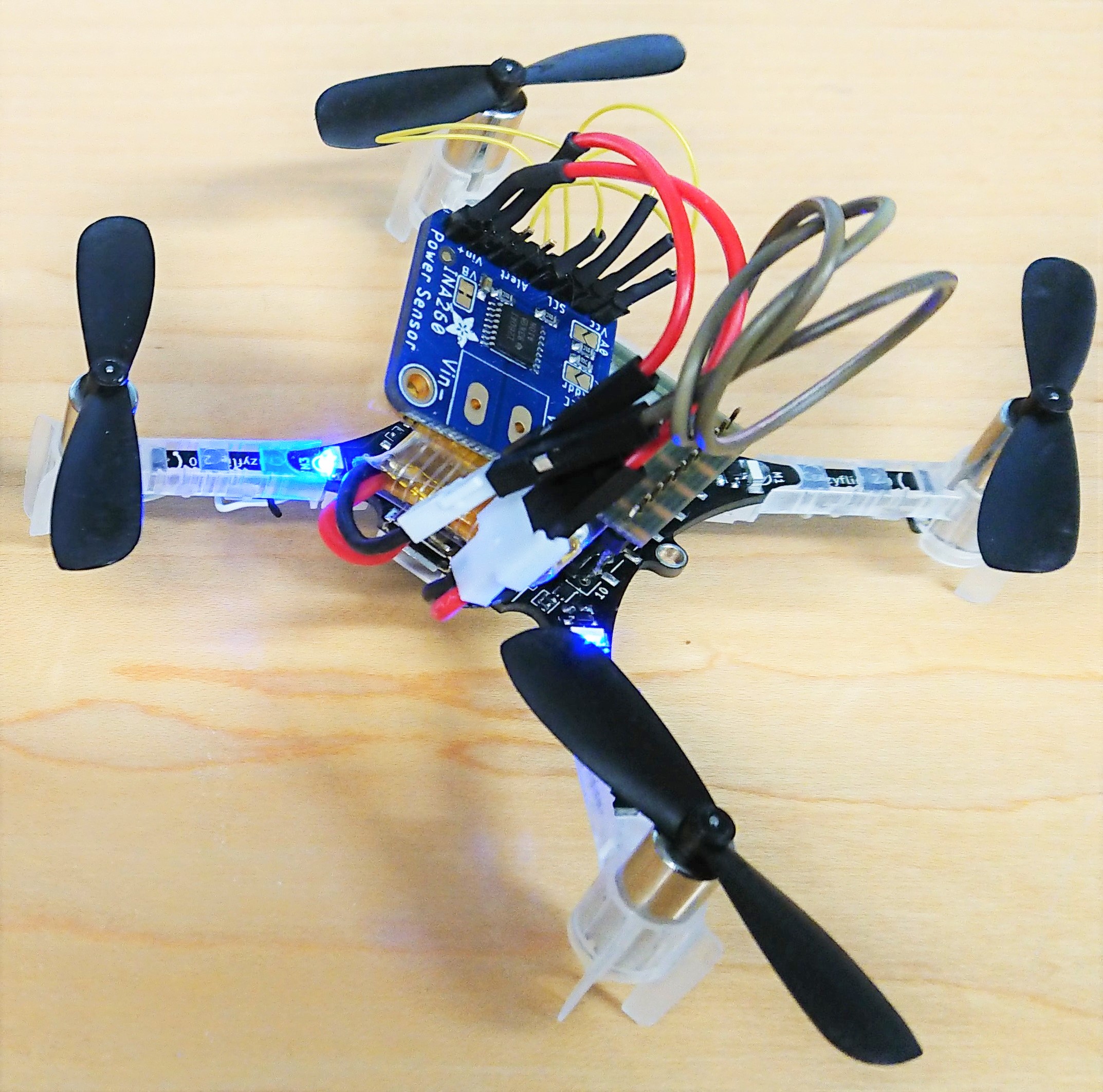}
                \caption{}
                \label{fig:cf-current-sensor}
        \end{subfigure}
        ~
		\begin{subfigure}[b]{0.45\linewidth}
       			\centering
                \includegraphics[width=\textwidth]{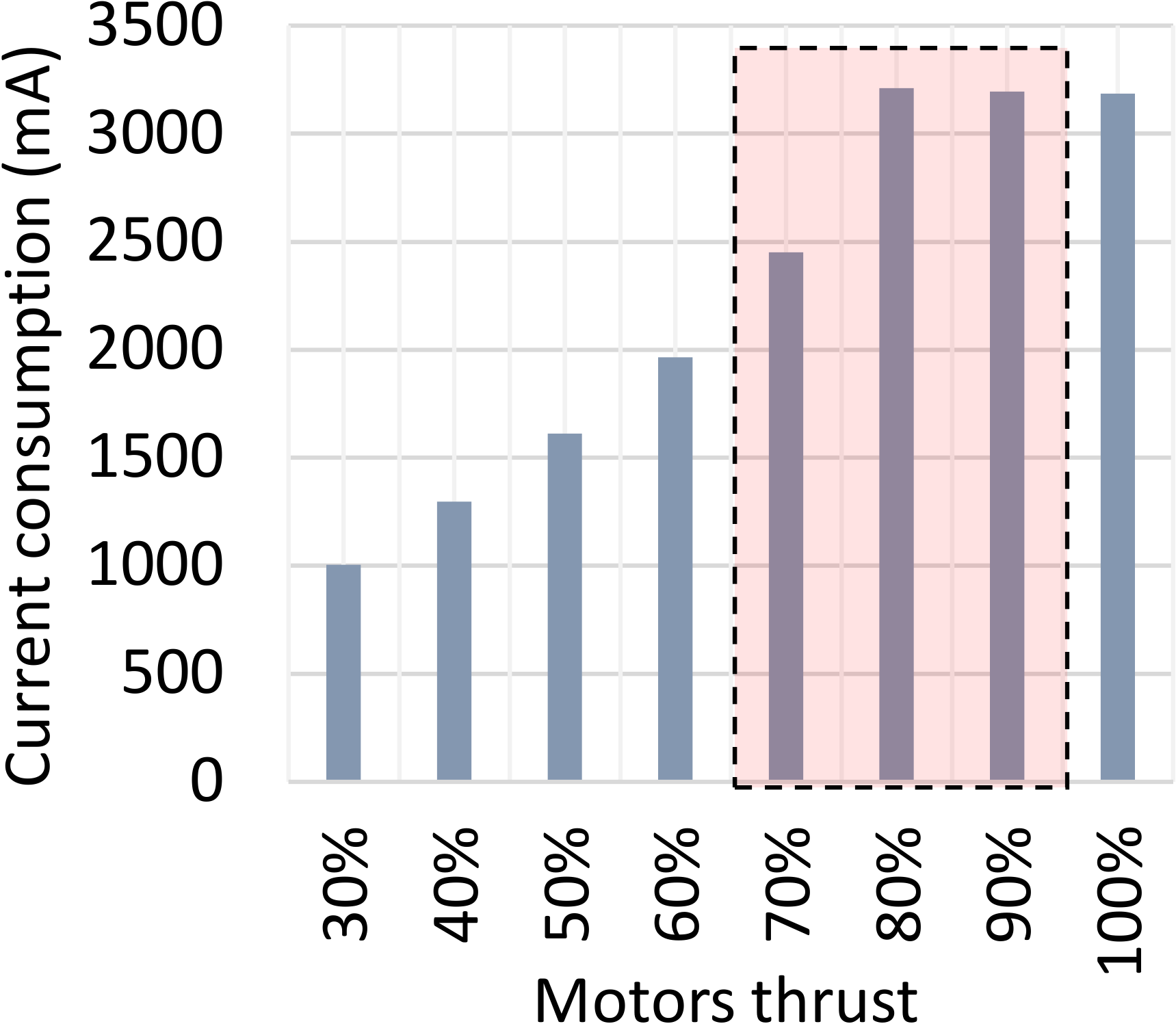}
                \caption{}
                \label{fig:cf-current-vs-thrust}
        \end{subfigure}
        }        
        \caption{(a) Crazyflie 2.0~\cite{giernacki2017crazyflie} equipped with TI INA260 current measurement sensor. 
        (b) \textcolor{debugcolor}{Crazyflie current consumption vs percentage of applied motor thrust. Annotated region shows
        the most frequently encountered virtual drone thrust percentage.}
		}
        \label{fig:met-deadlines-all-21}
\end{figure}

\textbf{Simulation traces:} The simulation toolchain comprises QGroundControl for waypoints visualization and uploading~\cite{meier2017qgroundcontrol}, 
PX4 autopilot~\cite{meier2015px4} as the flight controller of the simulated drone, and Gazebo as the drone dynamics simulator.
We use the Crazyflie 2.0 drone~\cite{giernacki2017crazyflie}, a completely open-source drone design supported by PX4 Autopilot 
and QGroundControl, for propulsion power dissipation profiling. %

Figure~\ref{fig:power-model} illustrates our methodology for deriving the propulsion power dissipation model of our target reference drone.
It consists of two parts, (i) the motor thrust profiling of configurable simulated drone instances (green colour) 
and (ii) the current consumption profiling of Crazyflie (peach colour).
Using the simulation part we \textcolor{debugcolor}{assembled a dataset of the required thrust per motor (expressed as a percentage of the maximum thrust of the simulated motors), for the
same flight trajectories under different payload mass and flying speed.
Using this dataset, we built piece-wise linear regression models to estimate the requested thrust per motor 
when the target drone flies between any new waypoints under certain payload mass and flying speed.
}

\textbf{Power measurements of Crazyflie drone:} The second part of our profiling methodology regards the mapping of the estimated motors' thrust to current consumption.
We retrofitted the Crazyflie 2.0 with a TI INA260 precision current monitoring circuit as shown in Figure~\ref{fig:cf-current-sensor}. 
This allows us to measure the current consumption of the Crazyflie, operating on battery supply, with 1\,mA granularity. %
Figure~\ref{fig:cf-current-vs-thrust} \textcolor{debugcolor}{shows this current consumption for different percentages of motors' thrust, applied to
all four motors simultaneously.
The annotated region of interest is the most frequently encountered thrust percentage of the simulated drones, when in flight.}

\textbf{Combining simulation traces with power measurements:} In total, our model combines the estimated thrust for the flight of the drone between two waypoints for a given payload and flight speed, with the current consumption model of the Crazyflie drone in order to produce a propulsion power consumption estimation.
The current limitation of our model is that Gazebo dynamics simulator does not support mini-drones and thus our analysis was
performed on heavier drones. 
To mitigate this weight difference, our estimations make use of the percentage of thrust recorded
from the simulated drones, instead of the absolute values.

%% file: sections/evaluation.tex
\section{Experimental Setup and Evaluation}
\label{sec:evaluation}

We automated the process of creating input test benchmarks with the use of random distributions
for the various target application parameters. 
We use the building locations of two real-world institutions (``A'' and ``C'') as the start depots.
We uniformly sample the area around the two institutions to create a parametric set of 5 to 10 target waypoints.
We use the same distribution to define the deadline of the waypoints, taking into
account the available speed range of the Crazyflie drone, which is set between 3 m/s and 8 m/s.

We evaluated a uniform distribution of payloads in three different weight ranges, representing \emph{light}, \emph{medium} and \emph{heavy} payload mass classes.
For each of the aforementioned parameters, we created three different benchmarks.
In the figures of the following Sections, \textit{we have labelled the benchmarks according to their payload weight
profile (L,M,H), depot location (``A'' or ``C''), and instance index.}
For example, \emph{M\_C3} represents the third medium payload mass benchmark, starting from the second institution location.
We performed the performance evaluation using PX4 source code, 
\textcolor{debugcolor}{for a target simulated drone of 1.0 kg without load, expected to fly along} the input benchmark waypoints.

\subsection{Comparative solution quality study}
\label{subsec:solution-efficiency}

First, we evaluate the ability of the \emph{greedy minimum distance algorithm} and our proposed \emph{simulated
annealing based algorithm} (Section~\ref{subsec:algorithms}) to compute trajectories which minimize
the number of missed deadlines. To allow us to compare our solutions against the optimal, we also performed exhaustive (brute force) evaluation of the possible deadline-minimizing trajectories. 
Because we generate our deadlines from a distribution, there are cases where even the optimal brute force approach cannot find a trajectory that meets all deadlines under a given
energy budget.
We use a battery capacity of 4000\,mAh as in the work of Baek et al.~\cite{baek2018battery}.
Each experiment corresponds to departure from the start depot, visiting of all target waypoints, and
successful return to the start depot.

Figure~\ref{fig:met-deadlines-all} shows the met deadlines of the calculated trajectories,
grouped according to the number of waypoints of each examined benchmark.
We observe that, for five waypoints (Figure~\ref{fig:met-deadlines-5Wps}) and six waypoints (Figure~\ref{fig:met-deadlines-6Wps}) our proposed solution is able to calculate a flight trajectory
which meets all deadlines and returns safely in the depot.
As target waypoints increase, the number of met deadlines decreases, but in all cases more than 70\% of the deadlines are met, averaging 98\% for 7 (Figure~\ref{fig:met-deadlines-7Wps}), 94\% for 8 (Figure~\ref{fig:met-deadlines-8Wps}), 
89\% for 9 (Figure~\ref{fig:met-deadlines-9Wps}) and 82\% for 10 (Figure~\ref{fig:met-deadlines-10Wps}) waypoints, respectively.
The number of met deadlines is frequently reduced when the average payload is heavy as the increased energy requirements 
must be compensated by decreasing the flight speed.

\subsection{Computational requirements analysis}

Apart from solution efficiency, an acceptable on-drone trajectory planning algorithm
must have constrained computational requirements given that (i) the drone hardware is bound by
limited computation resources and (ii) the trajectory planning module should not impose
a heavy latency overhead on the execution of drone software stack.

To achieve that, in the previous experiments the maximum search iterations of the 
SA optimizer were restricted to 5000.
We evaluate the required execution latency of the examined algorithms (excluding the brute-force one)
on two contemporary embedded Systems-on-Chip, which provide a reasonable tradeoff between computation efficiency and power consumption. 
First, an Intel Quark SoC with a single-core CPU at 400 MHz and 256 MB of RAM and second a dual-core 
ARM Cortex-A9 at 650 MHz with 512 MB of RAM memory, which has already been used in Aerotenna smart drone~\cite{skowron2019sense}.

\begin{figure}
       \makebox[\columnwidth]{
       \centering   
       \begin{subfigure}[b]{0.43\columnwidth}
       			\centering
                \includegraphics[width=\textwidth]{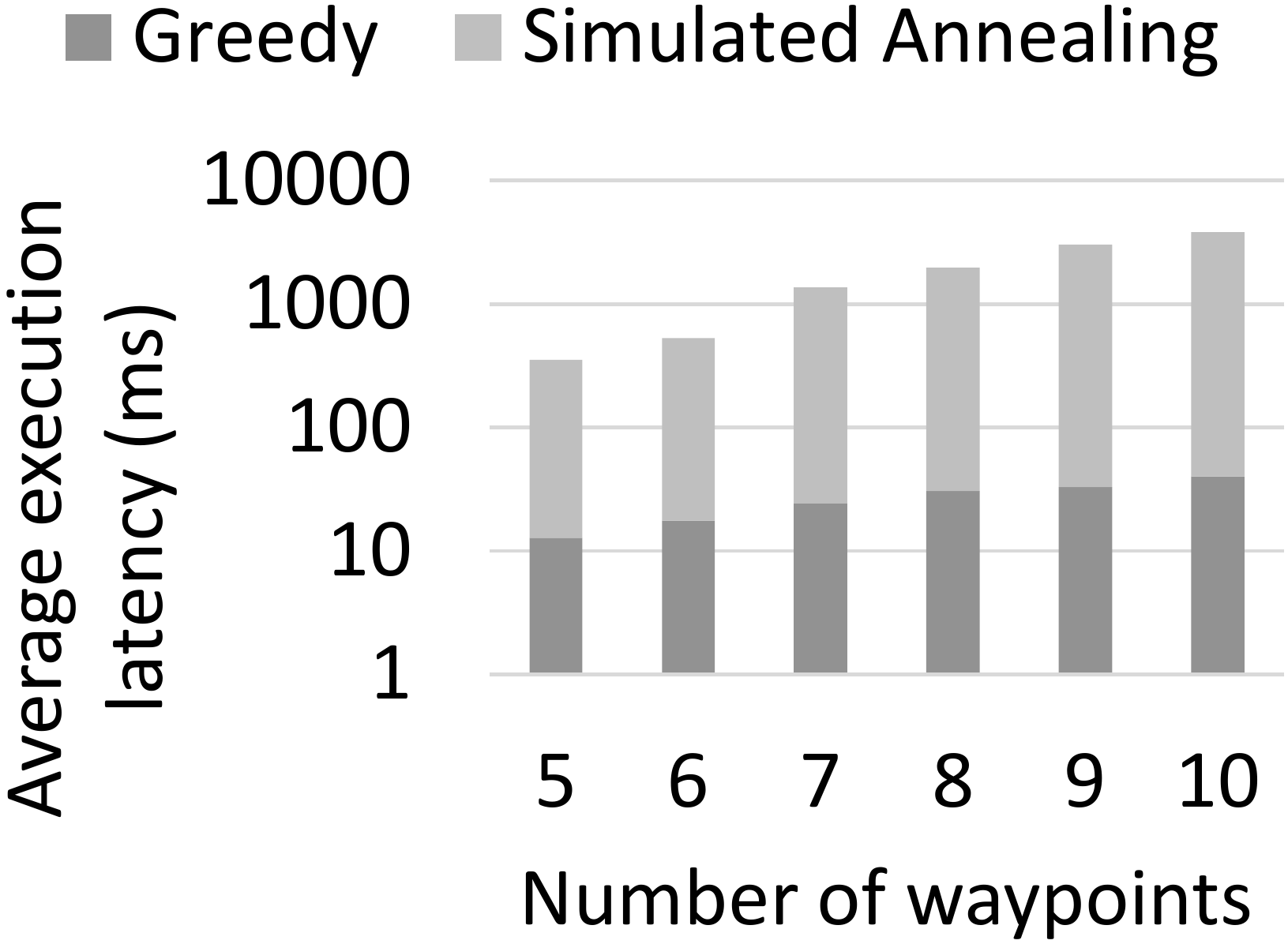}
                \caption{Intel Quark SoC.}
                \label{fig:avg-exec-latency-galileo}
        \end{subfigure}
        ~
		\begin{subfigure}[b]{0.43\columnwidth}
       			\centering
                \includegraphics[width=\textwidth]{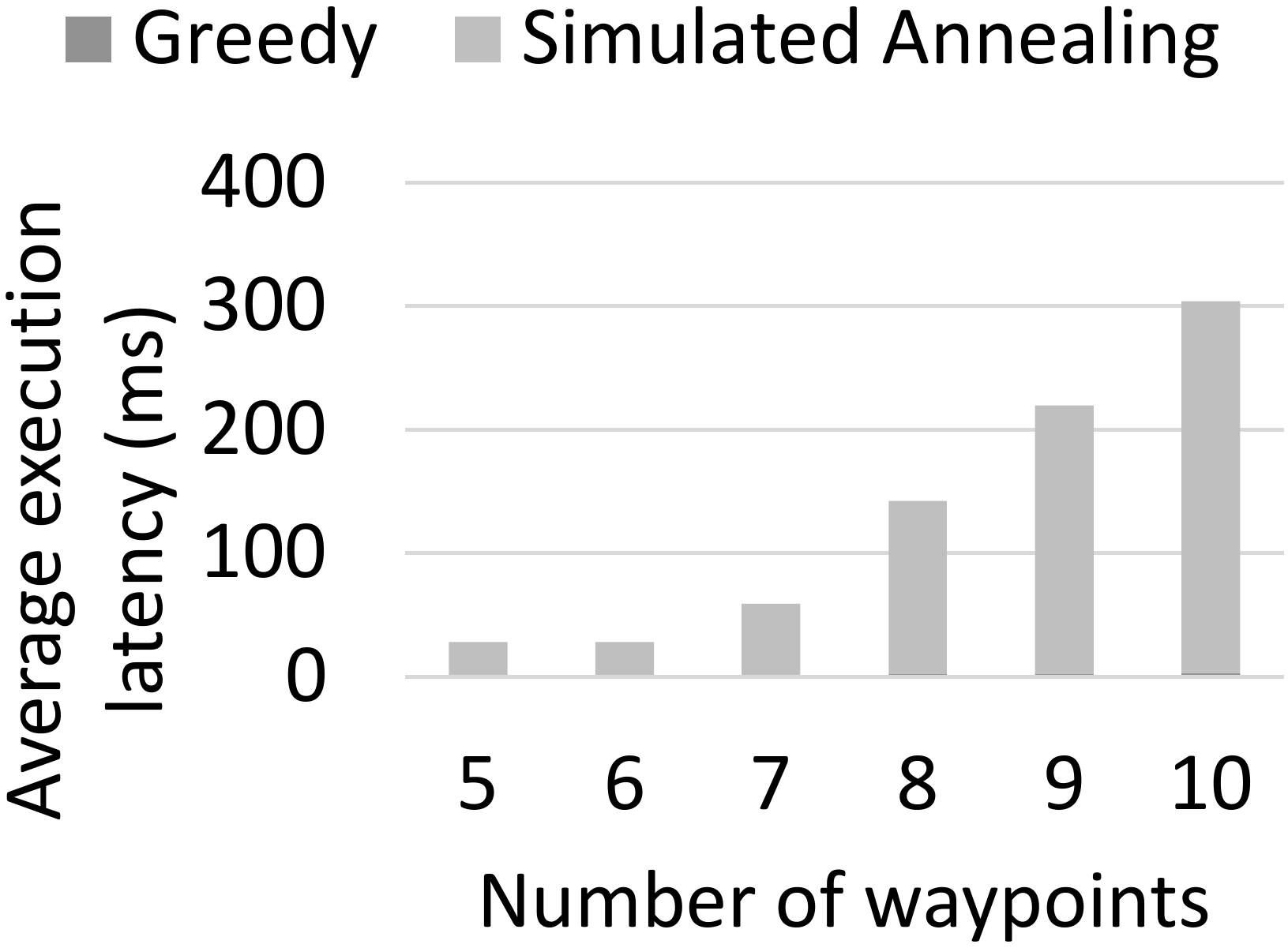}
                \caption{ARM Cortex-A9 SoC.}
                \label{fig:avg-exec-latency-ARM}
        \end{subfigure}
        }        
        \caption{Average measured execution latency of all benchmarks grouped according to number of visiting waypoints.}
        \label{fig:avg-exec-latency}
\end{figure}

Figure~\ref{fig:avg-exec-latency} summarizes the average execution latency of all examined benchmarks of Section~\ref{subsec:solution-efficiency},
on Intel Quark SoC (Figure~\ref{fig:avg-exec-latency-galileo}) and ARM Cortex-A9 (Figure~\ref{fig:avg-exec-latency-ARM}).
The required latency of the SA optimizer is stacked on top of the respective latency of the greedy heuristics,
given that their combination is the total execution latency of our proposed solution.
The latency is presented in millisecond granularity either in logarithmic scale in Figure~\ref{fig:avg-exec-latency-galileo}
or linear scale in Figure~\ref{fig:avg-exec-latency-ARM}.

In all cases, the required total latency is less than 4000\,ms, 
which is limited enough for on-drone trajectory planning.
In the Cortex-A9 SoC, the total latency always remained below 320\,ms, while the
latency of the greedy heuristic is always less than 10\,ms.
These values imply that on this system we can increase the search intensity and
achieve better results in terms of deadline miss minimization.

